\documentclass[12pt,preprint]{aastex}
\begin{document}

\title{PHYSICAL ORIGIN OF DIFFERENCES AMONG VARIOUS MEASURES OF SOLAR
MERIDIONAL CIRCULATION}
\author{MAUSUMI DIKPATI AND PETER A. GILMAN}
\affil{High Altitude Observatory, National Center for Atmospheric
Research \footnote{The National Center
for Atmospheric Research is sponsored by the
National Science Foundation. },
3080 Center Green, Boulder, Colorado 80301; dikpati@ucar.edu}
\author{ROGER K. ULRICH}
\affil{Department of Physics and Astronomy, University of California at Los
Angeles, Box 951547, Knudsen Hall, Los Angeles, California 90095-1547}

\begin{abstract}
We show that systematic differences between surface Doppler
and magnetic element tracking measures of solar meridional flow
can be explained by the effects of surface turbulent magnetic diffusion.
Feature-tracking speeds are lower than plasma speeds in low and 
mid-latitudes, because magnetic diffusion opposes poleward plasma flow
in low-latitudes whereas it adds to plasma flow at high latitudes. 
Flux transport dynamo models must input plasma flow; the model-outputs 
yield estimates of the surface magnetic feature tracking speed. We 
demonstrate that the differences between plasma speed and magnetic 
pattern speed in a flux-transport dynamo are consistent with the 
observed difference between these speeds.

\end{abstract}

\section{Introduction}
Observational estimates of meridional flow in the photosphere have been
made for many years using three principal methods. These include the 
tracking of surface features
such as magnetic elements \citep{sd96,hr10} and surface Doppler signal from
supergranulations \citep{sks06,sksb08}; Doppler shifts of 
selected spectral lines \citep{ulrich93,ub05}; most recently Doppler 
shifts of acoustic frequencies detected by helioseismic instruments 
\citep{ba10,gbs10}. Each of these methods has different strengths and weaknesses 
that have been discussed extensively in the published literature. They 
also are measuring different quantities, so they should not necessarily 
give the same results. In particular, both Doppler-based methods should 
be measuring the flow of the plasma, but the tracking methods are sensing 
the movements of features in the plasma, not necessarily the plasma itself.
By the laws of MHD, when magnetic diffusion is present, field lines and 
patterns can 'slip' relative to the plasma flow, since they are not 
'frozen in'. 

On the Sun we are dealing with turbulent magnetic diffusion, caused by 
the time changes in supergranulation \citep{leighton64}. Both
surface transport models and flux-transport dynamo models solve
equations that are in fact spatially and temporally averaged over
the detailed structure of supergranules; therefore both contain
turbulent diffusion, and in both models the ``mean'' field lines are 
not frozen in. There will also be magnetic diffusion on still smaller 
spatial scales, particularly within supergranules, but this effect should 
be isotropic \citep{sks06,sksb08} and not contribute to a net transport 
of flux in latitude. Outside of the immediate neighborhood of active 
regions, the turbulent magnetic diffusivity should be nearly homogeneous 
(independent of latitude and longitude) but magnetic flux will be diffused 
in latitude and longitude because in general there will be gradients of 
flux in both latitude and longitude. It is the difference between pattern 
speeds and plasma flow that we focus on in this paper.

It is well known that the surface of the Sun has a highly structured
magnetic field as well as turbulent motions. These turbulent motions
will also move patterns of magnetic features, in addition to
and in competition with the movement due to global flows such as
differential rotation and meridional circulation. The tracking of
magnetic features yields a 'flow' speed in latitude due to the combined
effects of meridional circulation and turbulence, so this speed is bound
to be different from that of the plasma flow. Is it possible to identify
and measure this difference, and learn about physics of the Sun from it?
What can flux transport dynamo models tell us about the difference
between pattern speeds and plasma speeds? Can the output of flux-transport
models estimate the turbulent diffusivity from the difference between
pattern speed and plasma flow? We attempt to answer these questions in the
following sections.

\section{Observational evidence of differences}

It is now possible to compare the measured meridional flow speeds from
surface Doppler, helioseismic and magnetic pattern tracking methods for
the whole of solar cycle 23. These measurements are displayed in Figure 1.
The red curve is the surface Doppler results obtained from the analysis 
of Mount Wilson Observatory data, already shown in
\citet{dgdu10}. The green curve is the helioseismic estimate from
\citet{ba10}, and the blue curve is for magnetic pattern tracking
reconstructed from the figure 3 of \citet{hr10}. Note that we omit
the uncertainty bars in each of the curves, because our focus is on
understanding the physical origin of one particular systematic difference
between the magnetic feature tracking-speed and Doppler-based plasma-speed,
rather than analyzing the accuracy of the methods. Also note that the
time-variation in the meridional flow within each solar cycle primarily 
arises due to the inflows toward the active regions \citet{zk04,skz07,
hht09}. But we have considered in Figure 1 the meridional flows averaged 
over the entire cycle, because we are focusing here on understanding the 
systematic difference between magnetic feature tracking speed and Doppler-based 
plasma-speeds.

We see that for the latitudes where both are measured, the surface Doppler
and helioseismic results are almost the same. Unfortunately helioseismic
methods can not reach to as high a latitude as can surface Doppler. We see
that the surface Doppler speed peaks around $20^{\circ}$, while the
helioseismic results peak at 30$^{\circ}$. Their peak amplitudes are within
$1\,{\rm m}\,{\rm s}^{-1}$ of each other. Both peaks are broad and it is not
clear that there is any difference in either peak amplitude or latitude
of the peak.

By contrast, the pattern tracking method yields a peak around $50^{\circ}$
latitude, which is 25-30 \% lower than the other peaks. The difference
between tracking and the other velocities is significant at all latitudes
equatorward of the peak. Poleward of its maximum, the differences are much
smaller. A difference between tracking-speed and plasma flow-speed of magnitude
$5\, {\rm m} {\rm s}^{-1}$ in the latitude range from $5^{\circ}$ up to about
$40^{\circ}$ magnitude have also been reported by \citet{skz07} and
\citet{szk07}. 

By heuristic reasoning, we show in the next section that the
differences between the pattern tracking and the other velocity measures
are due to the effects of magnetic diffusion on the pattern tracking speed.

\begin{figure}[hbt]
\epsscale{1.0}
\plotone{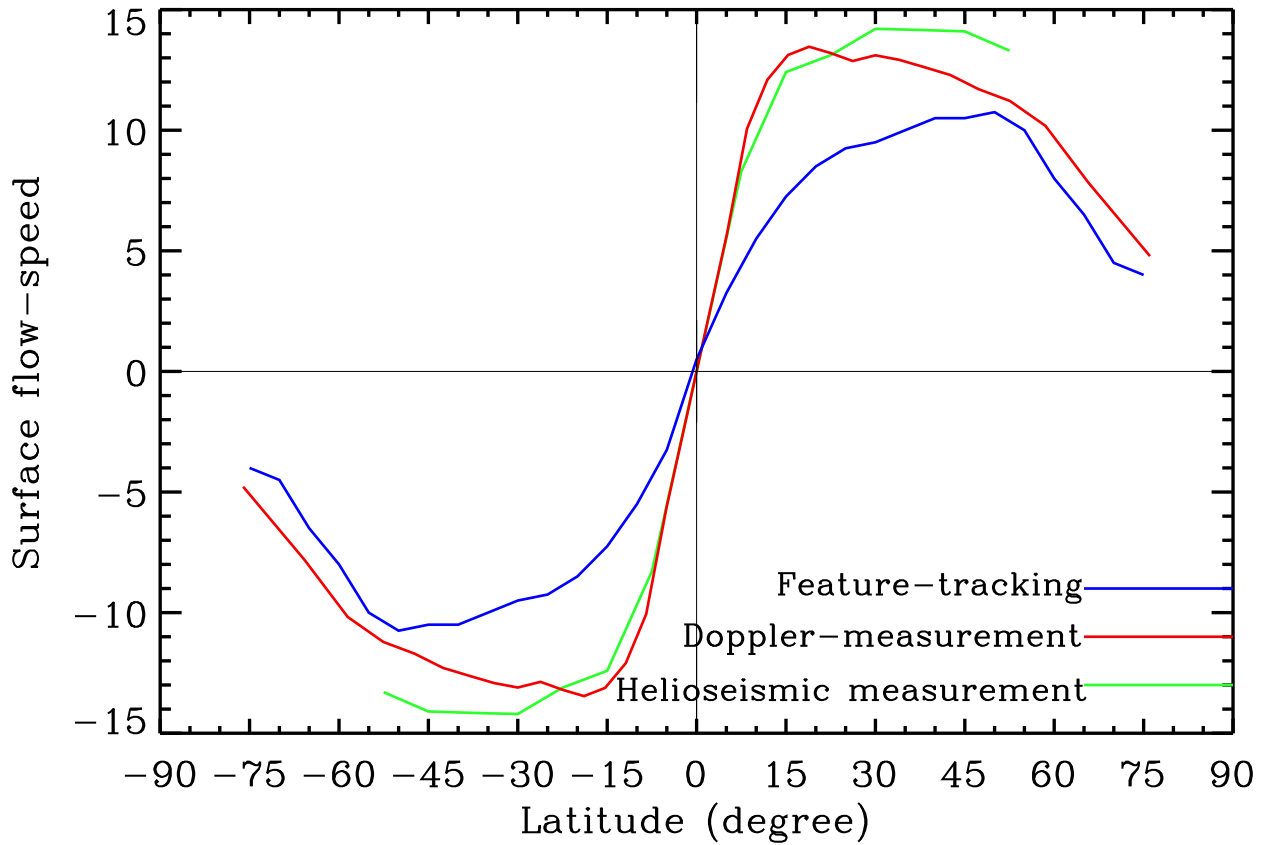}
\caption{Surface meridional flow speeds from various measures for solar 
cycle 23 (averaged over the entire cycle). Red curve: surface Doppler 
speed from Mount Wilson Observatory; 
green curve: helioseismic measurement from Basu \& Antia (2010); blue curve: 
magnetic feature movement read off of figure 3 of Hathaway \&
Rightmire (2010). }
\label{observed}
\end{figure}

Surface Doppler and magnetic tracking measures of meridional flow
also exist for cycle 22 \citep{dgdu10,ulrich93,sd96,latushko94,ch98}. For 
that period too the surface Doppler velocity is larger than the pattern 
velocity peak and occurs at a lower latitude. In addition, the plasma 
flow from surface Doppler measurements reverses sign and becomes equatorward 
above about $65^{\circ}$ latitude. This same behavior is seen in two 
tracking velocity profiles \citep{sd96,latushko94}. There too the reversed
tracking velocity is smaller than the surface Doppler velocity. This result
is also explained by the heuristic arguments we give below. 

\section{Heuristic explanation of differences in speed}

\begin{figure}[hbt]
\epsscale{1.0}
\plotone{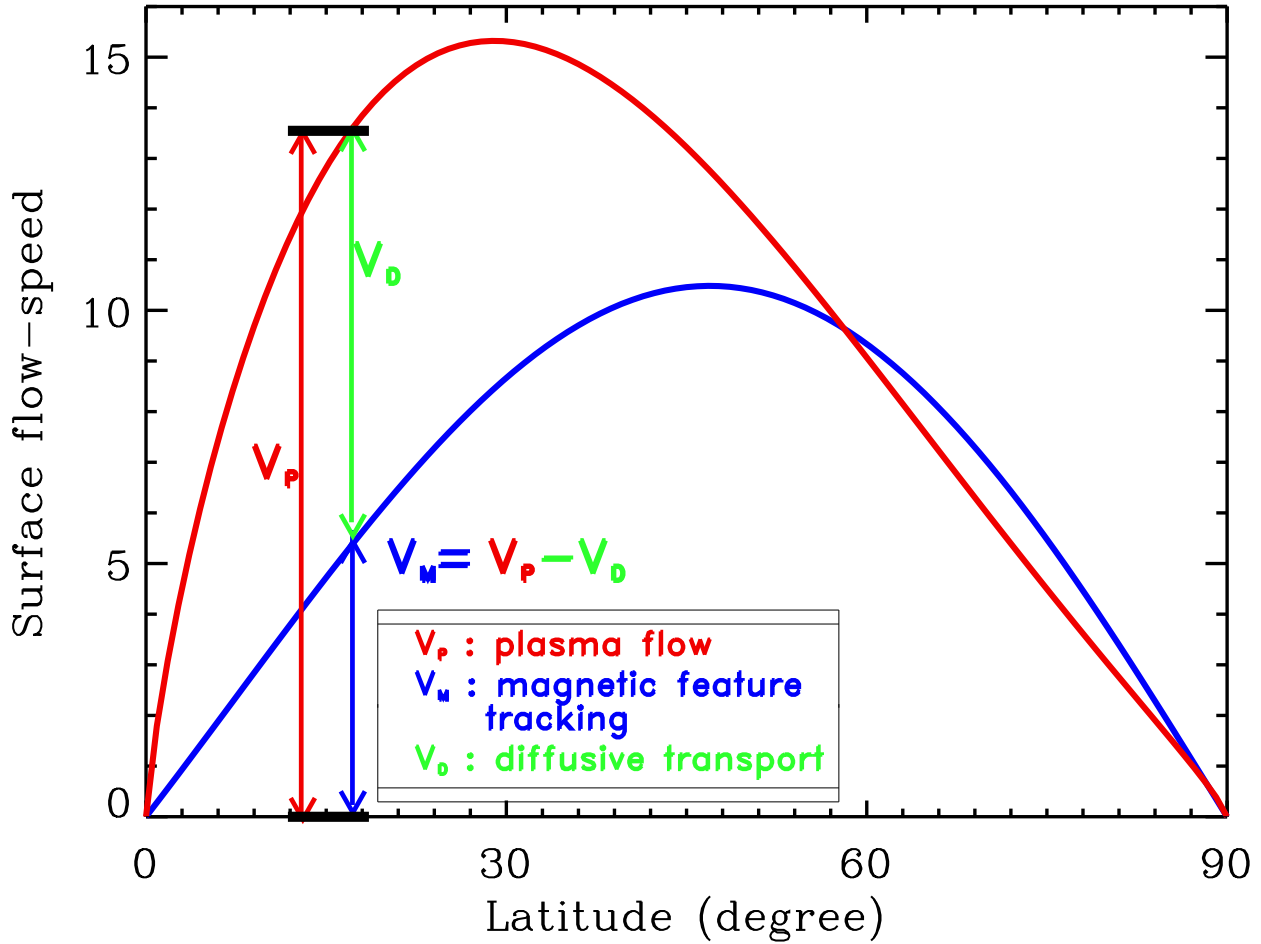}
\caption{Schematic of plasma (red curve and symbols) and magnetic feature 
tracking(blue curve and symbols) speeds, with difference due to diffusion 
speed (green arrow and symbols).}
\label{schematic}
\end{figure}

Starting from the observations of meridional velocities summarized
in the previous section, we give a heuristic physical explanation for
the systematic differences between the surface Doppler and
helioseismicly determined values, and those associated with the movement
of magnetic patterns. In schematic form, these differences are
illustrated in Figure 2. The red curve denotes a predominantly poleward
plasma velocity $V_P$ that peaks in the range $20^{\circ} - 30^{\circ}$
latitude.
The blue curve shows the magnetic pattern velocity $V_M$. It has a lower
peak amplitude compared to the plasma velocity and it peaks at a
substantially higher latitude, between $45^{\circ}$ and $55^{\circ}$.
The systematic difference between the two velocities is that at low
latitudes, the plasma velocity is faster, while at high latitudes the
two velocities are about the same.

These differences can be explained in terms of the turbulent magnetic
diffusion. We can define a 'diffusion velocity' $V_D$, given by
$\eta/L$, in which $\eta$ is the turbulent diffusivity, typically taken
to be $2-8 \times 10^{12} \, {\rm cm}{\rm s}^{-1}$ in the solar
photosphere, produced by mixing by supergranulation, and $L$
is a latitudinal length scale. $L$ should be some fraction of the
distance between equator and pole, a distance over which the
longitudinally averaged surface radial field varies by a substantial
amount. To illustrate the magnitude of this velocity, if we roughly take
$L=10^{10} \,{\rm cm}$ (about 1/10th of the distance between equator
and pole) we get $V_D=2-8 \, {\rm m}{\rm s}^{-1}$, values comparable
to the plasma flow.

Since the surface source of large scale surface poloidal fields is due
to the decay of emerged, tilted, bipolar active regions, the peak of the
source and hence of the surface radial fields is almost always found at
or near sunspot latitudes. In the absence of diffusion, these surface
poloidal fields would always be carried towards the poles by the
poleward plasma flow. We should expect an asymmetric apparent outflow 
from activity latitudes, because the turbulent diffusion
should increase the poleward transport rate in latitudes poleward of
spot latitudes, and decrease that transport rate in low latitudes. This
implies that the speed ($V_M$) of magnetic features should be reduced in
low latitudes, and increased poleward of sunspot latitudes. By this
reasoning, in low latitudes we should have $V_M=V_P-V_D$ while at high
latitudes we should have $V_M=V_P+V_D$. To a first approximation, $V_D$
represents the difference between the curves for $V_P$ and $V_M$.

In the immediate vicinity of active regions ($\sim 5^{\circ}$ latitude 
poleward and equatorward) there is an additional advective effect we 
have not included in the reasoning given above. This effect comes 
from the observation \citep{skz07,hht09} that there are persistent 
inflows of plasma from
both poleward and equatorward sides of active regions. These flows would
have the effect of reducing the apparent outflow of magnetic patterns.
This feature is relatively localized, however, and in this first study
we have not attempted to include it in the more global reasoning we
are focusing on here. It must be included in more advanced models
that examine in more detail the differences between plasma flow and 
magnetic feature tracking.

The observations in Figure 1 and the schematic in Figure 2 both
show the substantial difference between plasma and magnetic feature
tracking speeds occurs in low and mid latitudes, but not in high
latitudes. This implies that the diffusive 'speed' must be low there.
How is this possible? 
The concept of turbulent diffusion of magnetic elements across the 
solar surface due to the existence and evolution of supergranule patterns
originated with \citet{leighton64}. In magnetic regions supergranules are 
distorted and difficult to distinguish from the magnetic features,
because supergranules can be detected either in line-of-sight velocity
or Ca-K maps, in which magnetic features are also seen \citep{sks09}. 
Therefore, except in the immediate neighborhood of active regions, 
supergranule characteristics (horizontal scale, lifetime) are
seen to be nearly independent of latitude and longitude. It follows
that, for the diffusive speed to be much lower in high latitudes
than low, the gradient of the surface radial field must be on average
much lower there than equatorward of the peak in the poloidal source.
The diffusive transport rate is given by ${\eta \over r}{\partial B_r
\over \partial \theta}$. The diffusive speed associated with this
transport is given by $\eta/r\Delta \theta$, where $\Delta\theta$ is
the difference in latitude over which the difference in radial fields
is measured. The quantity $L=r\Delta \theta$ is, therefore, our length
scale.

\begin{figure}[hbt]
\epsscale{1.0}
\plotone{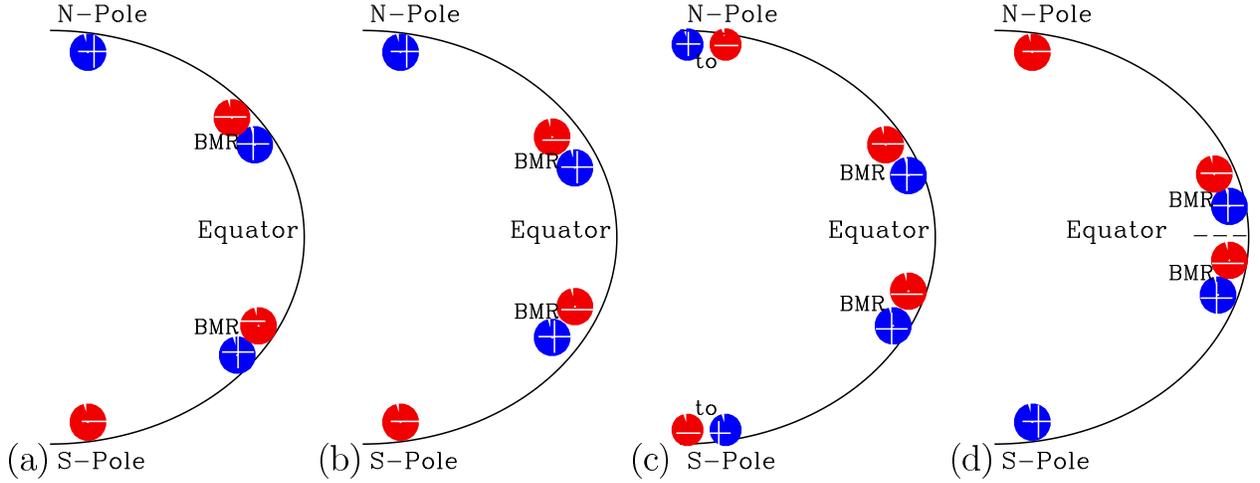}
\caption{Schematic of latitude locations of peaks in radial field
(positive peaks in blue, negative ones in red) for various sunspot cycle 
phases. a)cycle onset; b)ascending phase; c) maximum; d) descending phase. 
BMR stands for bipolar magnetic regions.}

\label{diffusive}
\end{figure}

We can explain the differences between the profiles of plasma flow
and magnetic pattern tracking speeds by considering qualitatively
the latitudinal locations of the strongest radial fields of both
polarities at four different phases of a solar cycle, namely at cycle
onset, ascending phase, maximum, and descending phases (see Figure 3). 
At cycle onset (frame 3a), peak radial fields of opposite polarities 
are at the poles and at the location of the follower fields of sunspots; 
peak radial fields are also at the locations of the leading polarity 
spots in North and South hemispheres. In this phase of the cycle, the 
latitudinal distance between follower polarity spots and the pole of 
that hemisphere is about $60^{\circ}$ in latitude. The distance between 
leader polarities in North and South hemispheres is also about $60^{\circ}$. 
So the latitudinal length scale for the evaluation of latitudinal gradients
in radial fields is about the same in the two cases. Since polar fields 
are usually weaker than leading polarity fields in low latitude, the 
gradient of radial field is somewhat smaller between the follower fields 
and the pole of the same hemisphere than between the leader polarities in 
North and South hemispheres. Therefore even at this phase, the diffusive 
velocity toward the equator should be somewhat larger than that toward 
the poles above sunspot latitudes.

This difference gets bigger as the cycle progresses. During the ascending 
phase (frame 3b), the distance between the follower polarities and their 
respective poles increases, while the distance between leader polarities 
in North and South decreases. So the low latitude gradients in radial 
fields increase relative to the high latitude gradients, with a corresponding 
increase in the diffusive speed, as defined earlier, in low latitudes 
compared to high latitudes. This trend continues through the maximum 
(frame 3c) and descending (frame 3d) phases. Therefore the difference in 
gradients of radial field that drives the diffusive transport continues 
to increase. In addition, at maximum, the polarity of the radial fields 
at the poles switches to become the same as that of the follower spot 
polarity in that hemisphere, while the polarities of leader spots in the 
two hemispheres remain opposite. This effect reduces the high latitude 
gradient relative to the low latitude gradient even more.

The net effect of these differences in gradient in radial fields, when
averaged over a whole solar cycle, is to make the diffusive velocity
directed toward the equator in low latitudes substantially larger than
the diffusive velocity directed toward the poles at active-regions'
latitudes and higher. Hence the magnetic pattern tracking velocity is
substantially lower than the plasma flow in low and mid latitudes 
compared to that at high latitudes, as seen in Figures 1 and 2.

Some analyses of magnetic pattern data \citep{sd96} indicate
the possibility of poleward movement between the equator and $10^{\circ}$
latitude, and also equatorward movement poleward of about $60^{\circ}$.
For simplicity, we have not included either of these feature in the
schematic. In the case of the poleward low latitude movement, this could
come about only if the turbulent diffusion were low enough, or the
poleward plasma flow large enough. If this pattern persisted for a
substantial fraction of a solar cycle, then during that cycle rather
little radial magnetic flux would be annihilated at the equator.

Equatorward high latitude magnetic pattern movement could only come
about if there is a second plasma meridional circulation cell there that
is sufficiently strong to prevent flux from traveling all the way to
the poles. If such a pattern persisted for most of a cycle, then the
conveyor belt would carry flux down to the bottom near the interface
of the two circulation cells rather than at the poles. This result is
qualitatively consistent with the surface Doppler estimates of
meridional flow by \citet{ub05}, and could lead to a shorter solar 
cycle, as discussed in \citet{dgdu10}.

\section{Pattern velocity from flux-transport dynamo solutions }

In this section we investigate the latitudinal movement of magnetic 
patterns in a flux transport dynamo, running in a weakly non-linear
kinematic regime. We simulate how a bipole source, inserted into the 
topmost layer (between $0.97R$ and $1R$) of the model, moves under the 
influence of a specified meridional flow single-cell as used in 
\citet{dgdu10} with a maximum flow-speed of 15 ${\rm m}\,{\rm s}^{-1}$ 
and turbulent magnetic diffusion of value $3\times 10^{12} \, {\rm cm}^2 
\, {\rm s}^{-1}$. The bipole source 
is a Gaussian in the poloidal potential with full width at half maximum of 
$6^{\circ}$. This source results in 'leader' and 'follower' radial fields 
with Maxwellian profiles, each having full width at half maximum of 
$3^{\circ}$ in latitude, with the Maxwellian 'tail' of the follower (leader) 
polarity fields extending toward the poles (equator). The source satisfies 
the condition that $\nabla \cdot B=0$. We solve the induction equation in the 
axisymmetric regime in two dimensional $r-\theta$ plane and study the 
poleward transport of radial fields after they have been induced through 
the deposition of bipoles.

\begin{figure}[hbt]
\epsscale{1.0}
\plotone{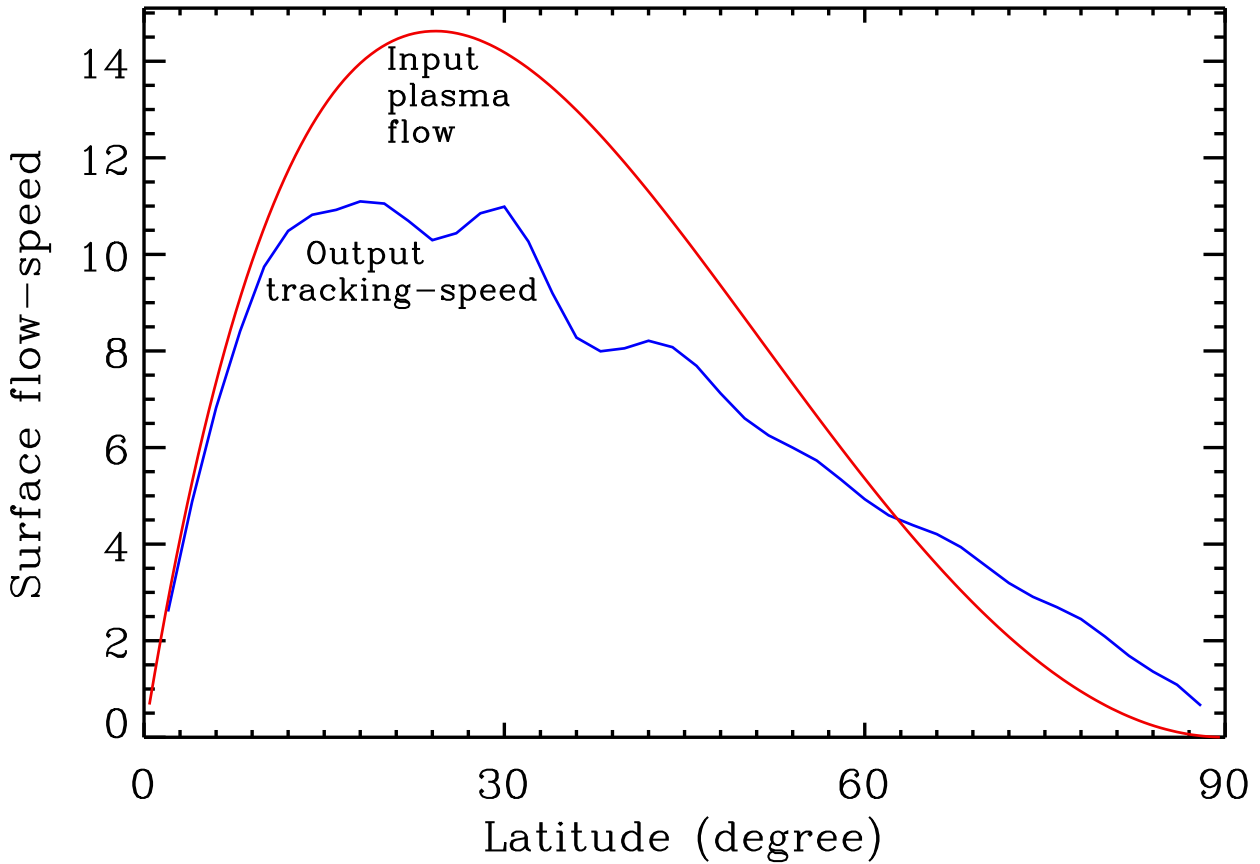}
\caption{Comparison of surface plasma flow (red curve) used as input to flux 
transport dynamo, and simulation-produced magnetic pattern tracking speed
(blue curve), derived from model-output.}

\label{simulation}
\end{figure}

We performed a series of numerical experiments in which the bipoles are
induced into the model for a period of 12 days, after which this source is
removed. All the physical processes in the model act on this source from
time t=0. After removal of the source, the latitudinal position of a radial
magnetic field contour of 30 Gauss of the follower polarity, which is on
approximately the central part the Gaussian in radial fields, is tracked.
From this tracking over a few weeks, a velocity is computed.

The experiments are performed on poloidal sources inserted every two
degrees of latitude between $2^{\circ}$ and $88^{\circ}$ latitude,
thereby generating a set of 44 velocities between equator and pole. These
velocities are plotted (blue curve) in Figure 4, in which the plasma
surface-flow speed that has been used as the input to the model is also
plotted (red curve). The result clearly shows that in low and
mid-latitudes,
the tracking speed is always lower than the plasma flow speed, by an amount
that is similar to that observed. Only poleward of $60^{\circ}$ latitude
does the tracking speed exceed that of the plasma flow. That occurs where
both speeds are low.

These experiments differ in some details from the heuristic
description given in section 3, for example in the fact that in these
experiments there are no polar fields to increase the magnitude of the
high latitude gradients and diffusive velocities. Therefore we should
expect to see the tracking velocity become larger than the plasma flow 
near the pole where plasma flow and diffusion velocity are always in the 
same direction. The simulations by \citet{wrs09} using surface transport
models with and without diffusion, showed that the slopes of the poleward
surges were steeper when both flow and diffusion were included, than in
the case when only flow was present (compare their figures 15 and 16).

Another important transport mechanism has not been considered in the present
simulation, namely the turbulent pumping. The latitudinal component of 
the mean turbulent pumping as computed from the magnetoconvection 
simulations \citet{osbr02,kkos06} is predominantly equatorward and
peaks around $15^{\circ}$ latitudes and with a magnitude of about 
$1\,{\rm m}{\rm s}^{-1}$. The influence of both the radial and latitudinal
turbulent pumping have been studied in great detail by \citet{gd08}. If
such pumping is included in simulating the magnetic feature-tracking speed,
even a larger difference between the input plasma speed and the 
model-derived tracking speed would occur at low latitudes. This might
produce even better agreement between the model-derived tracking-speed
and the observed values shown in Figure 1.  

The result in Figure 4 confirms that in dynamo models the velocity from
tracking of magnetic features should be considered as an output, not as a 
substitute for plasma flow speed which must be used as input to the model.
In fact, about 25 years ago it was first shown by NRL scientists 
\citep{ds87,wns89} that a meridional plasma flow peaking at low-latitudes
around $10^{\circ}$ from the equator is required to correctly explain 
the surface magnetic features' evolutionary properties. \citet{sl08} also
considered a variety of meridional flows peaking at different latitudes
and showed that, to obtain the best-fit of observed polar field pattern 
of cycle 23 with their model-output, the flow must peak around 
$15^{\circ}-20^{\circ}$ latitudes. With much better techniques for 
measuring meridional flow, we now know that the plasma flow indeed does 
peak at low-latitudes (see red and green curves in Figure 1).

\section{Concluding remarks}

We have shown using heuristic reasoning and numerical simulations that
the differences in surface meridional flow speed between the solar plasma 
and magnetic patterns can be explained as due to the effects of surface 
turbulent magnetic diffusion. All observational analysis referred to
in this paper are done with valid methods. We repeat that the purpose
of this paper was to understand the physical origin of the differences
between tracking-speed and Doppler-based plasma speed. The plasma speed
should be used as input into flux-transport dynamo models, whereas the
observed magnetic pattern speed should be compared with the output of 
such models. 

A very simplified case of a simulated feature-tracking speed derived
from a flux-transport dynamo model-output has been presented here. 
It captures the physical origin of the differences between the plasma 
flow incorporated into the model and the tracking-speed derived from
the evolving magnetic features in the model. 
However, there are many more aspects of the difference between plasma 
flow and magnetic pattern movement, which can be further explored using 
our model as well as other dynamo models benchmarked by \citet{jea08}.
Studies need to be done using models with higher magnetic diffusivity 
inside the convection zone \citep{ynm08}, with the magnetic diffusivity 
quenched due to the presence of strong magnetic fields \citep{gdd09}
and also with turbulent pumping included \citep{gd08}. Perhaps the 
importance of physical ingredients below the surface layer can be 
assessed by comparing model-derived tracking-speed from dynamo models 
with those obtained from surface-transport models \citep{wrs09,sl08}.  

From the input of Doppler plasma flow in all such models,
the comparison of model-derived tracking-speed with the observed
tracking-speed could give new estimates of the turbulent diffusivity 
on the surface of the Sun. Other plasma flow profiles could also be used, 
including ones that allow for a second, reversed plasma meridional flow cell 
in high latitudes as used by \citet{jb07,bebr05,dgdu10} to see how strong 
such a flow must be to reverse the 
magnetic pattern movement, as found for cycle 22 by \citet{sd96}. Different
representations of the bipole source could also be explored.

The simulation presented in \S4 was performed in an idealized environment, 
namely without any background magnetic fields present in the Sun. Reality
is more complex than that. For example, at high latitudes background polar 
fields are present with certain polarity and strength, and at sunspot latitudes
the active regions' fields appear and disappear as function of solar cycle.
A forthcoming paper will incorporate these complex background magnetic fields 
in the Sun in order to investigate further the differences between the 
input plasma flow and the output tracking-speed.
  
All of the flux-transport dynamo models discussed above are axisymmetric
and therefore only include the longitude-averaged meridional flow. Given
the recent observations of strong inflow-cells around active regions
\citep{skz07,hht09}, which
have not only latitude dependence but also longitude dependence, flux-transport
dynamos need to be generalized to include longitudinal dependence in
meridional flow.

\acknowledgements

We thank Sarbani Basu for supplying us their meridional flow data from 
their recent paper that is currently in press. We extend our thanks to 
Laurent Gizon for sharing his meridional flow data and for many helpful 
discussions. We also thank an anonymous reviewer for many helpful comments
which have helped improve the paper. This work is partially supported by 
NASA's Living With a Star program through the grant NNX08AQ34G. 


\end{document}